\documentclass[aps,floats,floatfix,nofootinbib,prl,superscriptaddress,tightenlines]{revtex4}
\usepackage{graphicx} \usepackage{bm} \usepackage{epsfig}
\usepackage{pstricks}

\newcommand{\beq}{\begin{equation}}
\newcommand{\eeq}{\end{equation}}
\newcommand{\bea}{\begin{eqnarray}}
\newcommand{\eea}{\end{eqnarray}}

\begin{document}
\title{Comment on `On the next-to-leading order gravitational spin(1)-spin(2) dynamics' by J. Steinhoff et al.}
\author{Rafael A. Porto}
\address{Physics Department, University of California, Santa Barbara\\
Santa Barbara, 93106 USA}
\author{Ira Z. Rothstein}
\address{Carnegie Mellon University
Dept. of Physics, \\
Pittsburgh PA  15213, USA}

\begin{abstract}
In this comment we explain the discrepancy found  between the results in arXiv:0712.1716v1 for the 3PN spin-spin potential and those previously derived in gr-qc/0604099. We point out that to compare one must include sub-leading lower order spin-orbit effects which contribute to the spin-spin potential
once one transforms to the PN frame. When these effects are included the results
in  arXiv:0712.1716v1 do indeed reproduce those found in  gr-qc/0604099.
\end{abstract}

\maketitle

In \cite{eih} we applied recently developed Effective Field Theory (EFT) techniques \cite{nrgr,nrgr2} to compute the next to leading order (NLO)  potential due to spin(1)-spin(2) interactions in the Newton-Wigner (NW) spin supplementarity condition (SSC).  In \cite{jan} the NLO spin(1)-spin(2) was also calculated. Given our claim that, up to 4PN order the equations of motion in the spin(1)-spin(2) sector can be obtained via the traditional Hamiltonian approach \cite{eih,nrgr5}, the authors then searched for the canonical transformation necessary to go between the results, finding that no such transformation exists. However, in order to  compare one must also consider spin-orbit effects, including a subleading one which
becomes a spin(1)spin(2)-orbit term once written in terms of the coordinate velocity in the post-Newtonian one.  In fact, once this new term is added,  the result in \cite{jan}  
do indeed reproduce the potential calculated in \cite{eih}.\\

The spin-orbit 1.5PN potential is given by \cite{nrgr2,nrgr5}
\begin{equation}
\label{so15pn}
V^{so}_{1.5PN}= \frac{G_Nm_2}{r^2}n^j\left(S^{j0}_1+S^{jk}_1(v_1^k-2v^k_2)\right)
+ 1 \leftrightarrow 2,
\end{equation}
and therefore, it depends on $S^{j0}$. As we emphasized before \cite{nrgr5}, in our formalism the spin variable lives in a locally flat frame, where $S^{ab}= S^{\mu\nu}e^a_\mu e^b_\nu$ and $e^a_\mu$ is a local tetrad field such that $e^a_\mu e^b_\nu = g_{\mu\nu}$ \cite{nrgr5}. The spin algebra is thus given by the standard $SO(1,3)$, and the spin gravity interaction takes the form, 
$-\frac{1}{2} \omega^{ab}_\mu S_{ab}$ 
with $\omega_\mu^{ab}$ the Ricci rotation coefficients. Expanding around flat space we have
$e^b_\mu = \delta^b_\mu + \frac{h_{\nu\mu}}{2}\delta^b_\nu + \ldots,$
which allows us to obtain the Feynman rules with which the potential is calculated \cite{nrgr2,eih,nrgr5}.\\ 

On the other hand, the NW SSC \cite{regge} written in terms of the local spin and the coordinate velocity, $v^j \equiv \frac{dx^j}{dt}$, then reads 
\beq
\label{onshell}
\label{nw}
S_1^{i0}= \frac{1}{2} S_1^{ij}v_1^j + \frac{1}{2} S_1^{ij}e^j_0({\vec x}_1) +\ldots =\frac{1}{2}({\vec v}_1\times{\vec S}_1)^i  + \frac{G_N}{2r^2}\left(({\vec n} \times {\vec S}_2)\times {\vec S}_1\right)^i +\ldots ,
\eeq 
where we have $e^j_0 = H^j_0/2 = \frac{G_N}{r^2}({\vec n} \times {\vec S}_2)^j + \ldots$, which follows from the one point function, $\langle H^j_0 \rangle$, using the leading order spin Feynman rule $\frac{i}{2} h_{0i,j}S^{ij}$ \cite{nrgr2,eih},  or by simple inspection of the Kerr metric in harmonic gauge. A similar term follows for particle 2.\\

In the EFT approach \cite{nrgr,nrgr2,eih,nrgr5} each vertex scales with different power of spin, and velocity, and therefore spin(1)-spin(2) interactions are those for which there is a spin tensor on each vertex. On the other hand spin-orbit terms are those for which the spin of one of the bodies interacts with the motion of the companion. As we can see from (\ref{so15pn}) and (\ref{nw}), to obtain the full $S_1S_2$ Hamiltonian up to  3PN we need to include a spin(1)spin(2)-orbit term given by
\beq
\frac{G^2_N}{2r^2}\left( m_2 n^i S_1^{ij}e^j_0({\vec x}_1) - m_1 n^iS_2^{ij}e^j_0({\vec x}_2)\right) =
\frac{G^2_NM}{2r^4} \left( ({\vec S}_1\times{\vec n})\cdot ({\vec n}\times{\vec S}_2)\right) =  \frac{G^2_NM}{2r^4} \left( {\vec S}_1\cdot{\vec n} {\vec S}_2\cdot{\vec n} - {\vec S}_1\cdot{\vec S}_2\right),
\eeq
where $M=m_1+m_2$, 
\newpage
and we thus get
\begin{eqnarray}
\label{ss}
V^{s1s2} &=& -\frac{G_N}{2r^3}\left[ \vec{S}_1\cdot
\vec{S}_2\left({3\over2}\vec{v}_1\cdot\vec{v}_2-3\vec{v}_1\cdot{\vec{n}}\vec{v}_2\cdot{\vec{n}}
-\left(\vec{v}_1^2+\vec{v}_2^2\right)\right)
-\vec{S}_1\cdot\vec{v}_1\vec{S}_2\cdot\vec{v}_2-\frac{3}{2}\vec{S}_1\cdot\vec{v}_2\vec{S}_2\cdot\vec{v}_1+
\vec{S}_1\cdot\vec{v}_2\vec{S}_2\cdot\vec{v}_2\right.\nonumber\\
&+& \vec{S}_2\cdot\vec{v}_1\vec{S}_1\cdot\vec{v}_1+
3\vec{S}_1\cdot\vec{n}\vec{S}_2\cdot\vec{n}
\left(\vec{v}_1\cdot\vec{v}_2+5\vec{v}_1\cdot\vec{n}\vec{v}_2\cdot\vec{n}\right)
-3\vec{S}_1\cdot\vec{v}_1\vec{S}_2\cdot\vec{n}\vec{v}_2\cdot\vec{n}-
3\vec{S}_2\cdot\vec{v}_2\vec{S}_1\cdot\vec{n}\vec{v}_1\cdot\vec{n}
\nonumber\\
&+& \left.
3(\vec{v}_2\times\vec{S}_1)\cdot\vec{n}(\vec{v}_2\times\vec{S}_2)\cdot\vec{n}
+3(\vec{v}_1\times\vec{S}_1)\cdot\vec{n}(\vec{v}_1\times\vec{S}_2)\cdot\vec{n}-
\frac{3}{2}(\vec{v}_1\times\vec{S}_1)\cdot\vec{n}(\vec{v}_2\times\vec{S}_2)\cdot\vec{n}\right.\nonumber\\
&-& \left.
6(\vec{v}_1\times\vec{S}_2)\cdot\vec{n}(\vec{v}_2\times\vec{S}_1)\cdot\vec{n}
\right]+\frac{G^2_N M}{2r^4}\left(5\vec{S}_1\cdot\vec{S}_2-17\vec{S}_1\cdot\vec{n}\vec{S}_2\cdot\vec{n}\right) - \frac{G_N}{r^3}\left(\vec{S}_1\cdot\vec{S}_2-3\vec{S}_1\cdot\vec{n}\vec{S}_2\cdot\vec{n}\right),
\end{eqnarray}
where we also added last, the LO spin(1)-spin(2) potential at 2PN order.\\

We can now show that the canonical transformation generated by ($b=c=1/2$ in \cite{jan})
\beq
\label{g}
g = \frac{G_N}{2r^2} \left[ \frac{1}{m_2} {\vec S}_1\cdot {\vec p}_2 {\vec S}_2\cdot {\vec n} -\frac{1}{m_1}{\vec S}_2\cdot {\vec p}_1 {\vec S}_1\cdot {\vec n} \right] +  \frac{G_N}{2r^2} \left[ \frac{1}{m_1} {\vec S}_1\cdot {\vec S}_2 {\vec p}_1\cdot {\vec n} -\frac{1}{m_2}{\vec S}_2\cdot {\vec S}_1 {\vec p}_2\cdot {\vec n} \right], 
\eeq
 leads to  $\delta H^{NLO}_{SS}=0$ at each order in $G_N$.  
Thus the results in \cite{jan} do indeed reproduce those in \cite{eih}.  In a forthcoming paper we will present full details of the spin-spin calculation.
\\

\end{document}